\documentclass[12pt]{article}
\usepackage{amsmath,amsfonts,amssymb}
\usepackage{graphicx}
\usepackage{enumerate}
\usepackage{natbib}

\usepackage{float}
\usepackage[caption = false]{subfig}
\usepackage{caption}
\usepackage[flushleft]{threeparttable}
\usepackage{booktabs}
\usepackage{color}
\usepackage[breaklinks=true]{hyperref}
\hypersetup{
	colorlinks   = true,
	citecolor    =  blue
}
\usepackage{bm}
\newcommand{\blind}{1}

\newtheorem{thm}{Theorem}

\newtheorem{prop}{Proposition}
\newtheorem{ass}{Assumption}

\newtheorem{rem}{Remark}

\DeclareMathOperator*{\argmin}{arg\min}
\DeclareMathOperator*{\argmax}{arg\max}

\begin{document}

\def\spacingset#1{\renewcommand{\baselinestretch}%
{#1}\small\normalsize} \spacingset{1}


\if1\blind
{
  \title{\bf Estimation of Characteristics-based Quantile Factor Models\thanks{
    We are highly indebted to Yuan Liao for extremely useful comments. We also thank participants at the 2nd Workshop on  High Dimensional Data Analysis. Financial support from MICIN/AEI/10.13039/501100011033, grants PID2019-104960GB-I00, PID2020-118659RB-I00, TED2021-129784BI00, CEX2021-001181-M, and Comunidad de Madrid, grants EPUC3M11 (V PRICIT) and H2019/HUM-5891, is gratefully acknowledged. The usual disclaimer applies.}}
  \author{Liang Chen\hspace{.2cm}\\
    HSBC Business School, Peking University\\
    and \\
    Juan J. Dolado \\
    Department of Economics, Universidad Carlos III de Madrid \\
    and \\
    Jes\'{u}s Gonzalo \\
    Department of Economics, Universidad Carlos III de Madrid \\
    and \\
    Haozi Pan \\
    HSBC Business School, Peking University 
    }
  \maketitle
} \fi

\if0\blind
{
  \bigskip
  \bigskip
  \bigskip
  \begin{center}
    {\LARGE\bf Estimation of Characteristics-based Quantile Factor Models}
\end{center}
  \medskip
} \fi

\bigskip
\begin{abstract}
This paper studies the estimation of characteristic-based quantile factor models where the factor loadings are unknown functions of observed individual characteristics while the idiosyncratic error terms are subject to conditional quantile restrictions. We propose a three-stage estimation procedure that is easily implementable in practice and has nice properties. The convergence rates, the limiting distributions of the estimated factors and loading functions, and a consistent selection criterion for the number of factors at each quantile are derived under general conditions. The proposed estimation methodology is shown to work satisfactorily when: (i) the idiosyncratic errors have heavy tails, (ii) the time dimension of the panel dataset is not large, and (iii) the number of factors exceeds the number of characteristics. Finite sample simulations and an empirical application aimed at estimating the loading functions of the daily returns of a large panel of S\&P500 index securities help illustrate these properties.   
\end{abstract}

\noindent%
\vfill

\newpage
\spacingset{1.9} 

\section{Introduction}
The generalization of the classical factor analysis has taken place through two main approaches. On the one hand, there has been the development of approximate factor models (\textbf{AFM}) where factors are assumed to be unobserved and therefore need to be jointly estimated with their loadings.\footnote{AFM were first proposed by \cite{10.2307/1912275} to characterize the co-movement of a large set of financial asset returns. The estimation and inference theory of these models and their subsequent extensions have been developed, \textit{inter alia}, by \cite{stock2002forecasting}, \cite{bai2002determining}, and \cite{bai2003inferential}; see \cite*{fan2021recent} for a recent overview of this line of research.}  As a result, AFM suffer from a generic identification problem since both sets of objects can only be identified up to a rotation matrix.  On the other hand, a growing literature has emerged in finance trying to explain the cross-sectional co-movements of stock returns on the basis of observable factors. In this setup, the factors are usually approximated using the differences between the returns of portfolios sorted by some observed characteristics, e.g. market capitalization and book-to-market ratio. This popular approach, pioneered by \cite{fama1993common}, has been extended to include some additional factors, such as \textit{momentum, profitability} and \textit{investment}, together with the well-known Fama-French three factors (see \cite{fama2015five}). 

Both approaches have pros and cons. In effect, while the \textit{latent factors} approach relies on  easily implementable estimation methods, such as  the \textit{principal component analysis} (\textbf{PCA}), it is often criticized for the lack of interpretation of the estimated factors. Conversely, the Fama-French approach turns out to be unambiguous about this interpretation; yet, their method of constructing the factor proxies quickly becomes unreliable for typical sample sizes as the number of factors grows (see \cite{connor2007semiparametric}). 

A setup that tries to take advantage of both approaches while avoiding their shortcomings is the so-called \textit{characteristic-based factor models} (\textbf{CFM}) introduced by \cite{connor2007semiparametric} and later extended by \cite*{connor2012efficient}. According to this framework,  the factor loadings are assumed to be smooth nonlinear functions of some observed characteristics of the different units, while the factors remain unobserved, as in AFM. Thereby, the latent factors in CFM can be easily estimated even when the number of factors is not small, whereas their interpretation hinges on the choice of the  observed characteristics. Subsequently, \cite*{fan2016projected} have generalized this framework by allowing both the number of factors to differ from the number of observed characteristics and the factor loadings not to be fully explained by those characteristics. For the estimation of this kind of models, these authors introduced a new methodology called \textit{projected principal component analysis} (\textbf{PPCA}), showing that such estimators exhibit faster rates of convergence than the conventional PCA estimators for AFM.

The main goal in this paper is to extend the analysis of \cite{connor2012efficient} and \cite{fan2016projected} to a new class of factor models labeled \textit{characteristic-based quantile factor models} (\textbf{CQFM}). In relation to CFM, the main difference is that the idiosyncratic errors in CQFM are subject to quantile restrictions instead of mean restrictions. Moreover, in CQFM the latent factors, the loading functions, and the number of factors are all allowed to vary across quantiles, providing in this fashion a more complete picture of how the joint distributions of many asset returns are driven by a few common risk factors.

In particular, our main contribution here is to provide a new three-stage estimation method for CQFM, labeled \textit{quantile-projected principal components analysis} (\textbf{QPPCA}), which is computationally simpler than the other available estimation procedures for this kind of factor models. This procedure works as follows. First, at each time period, the outcomes (e.g. stock returns) are projected onto the space of the observed characteristics by means of sieve quantile regressions. Second, using the fitted values from the first step, their factors and loadings are estimated by means of PCA. Finally, the whole loading functions are retrieved by projecting the estimated loadings onto the basis of the sieve space. In addition, we propose a novel estimator for the number of factors at each quantile which is shown to perform satisfactorily for reasonable sample sizes.      

The rates of convergence and the limiting distributions of the estimated factors and loading functions in QPPCA are derived under very general conditions, whereas the estimator for the number of factors at each quantile is also shown to be consistent. In particular, a relevant contribution of QPPCA to this literature is that all its  asymptotic properties are obtained without assuming any moment restrictions on the idiosyncratic errors. Thus, it becomes a nice tool to analyze data from financial markets where the error distributions are known to have heavy tails. Moreover, we only require the number of cross section observations (denoted as $n$) to diverge in our asymptotic analysis, while the number of time series observations (denoted as $T$) can be taken to be either fixed or diverging. 

It is noteworthy that the QPPCA estimation method exhibits several similarities with the PPCA approach of \cite{fan2016projected}. Yet, the main difference is that, given our less restrictive assumption on the error terms, sieve quantile regressions are implemented in the first step to project the observed outcomes, whereas \cite{fan2016projected} uses sieve least square regressions. Thereby, QPPCA estimation turns out to be more robust to heavy tails and outliers, while the consistency of the PPCA estimators requires much stronger moment restrictions on the disturbance terms. However, the robustness of QPPCA estimators entails two potential costs: (i) the average convergence rate of the estimated factors is generally slower than that of the PPCA estimator, unless the strong assumption that the idiosyncratic errors are independent of the observed characteristics is made; and (ii) unlike PPCA, it has to be assumed that the factor loadings are fully explained by the observed characteristics (see below). 

A closely related paper to ours is \cite*{ma2021estimation}, which also addresses the estimation and inference for CQFM by considering a \textit{semiparametric quantile factor analysis} (\textbf{SQFA}) approach where observed characteristics are potentially allowed to affect stock returns in a nonlinear fashion. Thus, as QPPCA, SQFA extends \cite{connor2012efficient} to the quantile restriction case. Yet, several important differences exist between the two approaches. First, as in  \cite{connor2007semiparametric} and \cite{connor2012efficient}, SQFA assumes that the number of factors is known and is equal to the number of observed characteristics. In contrast, QPPCA does not only allow the number of factors (which can vary across quantiles) to be different from the number of characteristics, but also implies that the number of factors at each quantile can be consistently estimated from the data. Second, while SQFA's initial estimator of the quantile loading functions also relies on sieve quantile regressions, its subsequent steps are based on an iterative minimization algorithm to jointly estimate the factors and loadings. This algorithm can be computationally costly. This potential problem is easily solved with the QPPCA methodology since its second and third steps are based on PCA, which are much easier to compute. Third, the asymptotic results of \cite{ma2021estimation} are obtained as $n,T\rightarrow\infty$, while all our results hold either $T$ is fixed or $T\rightarrow\infty$ as $n\rightarrow\infty$. Additionally, there are some differences in the assumptions imposed in these two approaches, which will be further discussed in the next sections, once the main theoretical results are presented.

Next, it is important to highlight that the CQFM can be viewed as closely related to the quantile factor models (\textbf{QFM}) recently proposed by \cite*{chen2021quantile} to generalize AFM to quantile regressions. In effect, while no restrictions on the factor loadings are imposed in QFM (except for a standard rank condition), the loadings in CQFM are modeled as unknown functions of some observed characteristics to help interpret the latent factors though, like in SQFA, it also entails the risk of misspecification in the choice of the relevant characteristics. Furthermore, in relation to the quantile factor analysis (\textbf{QFA}) estimators proposed by \cite{chen2021quantile}, an additional advantage of the CQFM setup is that the model can be consistently estimated even when $T$ is fixed, while the QFA estimators are only consistent when both $n$ and $T$ go to infinity.  

Lastly, we provide an empirical application of the proposed estimators to analyze the behavior of the risk factors and their loadings in a panel dataset of excess stock returns that has been used in other studies. Our main finding is that the use of QPPCA allows for uncovering substantial variations of the estimated loading functions across different quantiles which cannot be obtained using PPCA. 

The outline of the rest of the paper is as follows. Section 2 introduces the model and the estimators. Section 3 derives the asymptotic properties of the proposed estimators and presents a novel consistent estimator of the number of factors at each quantile. Section 4 provides several Monte Carlo simulation results for finite samples. Section 5 is devoted to an empirical application of the proposed estimators. Finally, Section 6 concludes. An online appendix gathers detailed proofs of the theorems and specific results regarding several Monte Carlo simulations.

\vspace{0.3cm}
\noindent \textbf{Notations:}
For any matrix $\bm{C}$, $\|\bm{C}\|$ and $\|\bm{C}\|_S$ denote the Frobenius norm and the spectral norm of $\bm{C}$, respectively; $\lambda_{\min}$ and $\lambda_{\max}$ denote the minimum and maximum eigenvalues of $\bm{C}$, respectively, when the all eigenvalues are real; and $\bm{C}>0$ signifies that $\bm{C}$ is a positive definite matrix. For two sequences of positive constants $\{a_1,\ldots,a_n,\ldots\}$ and $\{b_1,\ldots,b_n,\ldots\}$, $a_n\asymp b_n$ means that $a_n/b_n$ is bounded below and above for all large $n$. The symbol $\lesssim$ means that the left side is bounded by a positive constant times the right side. Finally, for a random vector $(Y,X)$, $\mathsf{Q}_{\tau}[Y|X=x]$ denotes the $\tau$-quantile of $Y$ given $X=x$.

\vspace{0.3cm}
\noindent \textbf{Acronyms:}
Given the large number of acronyms used throughout the paper, we repeat them here (in the same order as they appear in the main text) to facilitate the reading of the paper: \textbf{AFM} (approximate factor model), \textbf{CFM} (characteristics-based factor model), \textbf{PPCA} (projected principal component analysis), \textbf{CQFM} (characteristics-based quantile factor model), \textbf{QPPCA}  (quantile-projected principal component analysis),  
 \textbf{SQFA} (semiparametric quantile factor analysis), \textbf{QFM} (quantile factor model), and \textbf{QFA} (quantile factor analysis).

 
\section{Model and Estimators}
\subsection{Model}
For a panel of observed data $\{y_{it}\}_{1\leq i \leq n, 1\leq t\leq T}$, \cite{chen2021quantile} consider the following quantile factor model (QFM): 
\begin{equation}\label{eq1}
y_{it} = \bm{\lambda}_i' (\tau)\bm{f}_t(\tau)+u_{it}(\tau), \quad \tau\in(0,1),
\end{equation}
where $\bm{\lambda}_i(\tau), \bm{f}_t(\tau) \in \mathbb{R}^{R}$ are quantile-dependent \textit{unobserved} quantile factor loadings and quantile factors, respectively, $R$ is the number of factors at quantile $\tau$, and $u_{it}(\tau)$ is the idiosyncratic error satisfying $\mathsf{Q}_{\tau}[ u_{it}(\tau)| \bm{\lambda}_i(\tau), \bm{f}_t(\tau)]=0$.\footnote{Note that the dependence of $R$ on $\tau$ is suppressed to ease the notations.} 

Our focus in this paper is on the CQFM model considered by \cite{ma2021estimation}, which can be viewed as nesting the special case of the QFM in \eqref{eq1} where $\bm{\lambda}_i(\tau)$ are unrestricted. In particular, let us assume the existence of a vector of \textit{observed} characteristics $\bm{x}_i=(x_{i1},x_{i2},..,x_{iD})\in\mathbb{R}^D$ for unit $i$ such that 
\begin{equation}\label{eq2} \bm{\lambda}_i(\tau)  = \bm{g}_{\tau}(\bm{x}_i), \end{equation}
where $\bm{g}_{\tau}(\cdot): \mathbb{R}^D \mapsto  \mathbb{R}^{R}$ is a vector of unknown functions for each $\tau$. As in \cite{connor2007semiparametric}, \cite{connor2012efficient} and \cite{fan2016projected}, we suppose that the $r$th element of $\bm{g}_{\tau}(\bm{x}_i)$ is given by the following additive function
\[ g_{\tau,r}(\bm{x}_i) = \sum_{d=1}^{D}  g_{\tau,rd}(x_{id}), \]
where $g_{\tau,r1},\ldots, g_{\tau,rD}$ are unknown functions. As in the related literature, it is assumed that $g_{\tau,r}$ is time-invariant so that the loadings capture the cross-sectional heterogeneity only. As \cite{fan2016projected} argued, such a specification is not stringent since in many factor-model applications to stationary time series, the analysis is carried out within each fixed time window with either a fixed or slowly-growing $T$. Yet, even if there are individual characteristics that are time-variant, like e.g. firm size or firm age, following these authors, we expect the conclusions in the current paper to remain valid if some smoothness assumptions are added for the time-varying components of those covariates.

Let $\bm{Y}$ be the $n\times T$ matrix of $y_{it}$, $\bm{F}_\tau$ be the $T\times R$ matrix of $\bm{f}_t(\tau)$, $\bm{X}$ be the $n\times D$ matrix of $\bm{x}_i$, $\bm{G}_{\tau}(\bm{X})$ be the $n\times R$ matrix of $\bm{g}_{\tau}(\bm{x}_i)$, $\bm{U}_{\tau}$ be the $n\times T$ matrix of $u_{it}(\tau)$. Then, models \eqref{eq1} and \eqref{eq2} can be rewritten in compact matrix form as:
\begin{equation}\label{eq3}
\bm{Y} = \bm{G}_{\tau}(\bm{X})\bm{F}_\tau'+\bm{U}_{\tau}.
\end{equation}

As already mentioned, the above setup is more general than those considered in the models of \cite{connor2012efficient} and \cite{ma2021estimation}. In the latter, the dimension of the vector of characteristics is required to be equal to the number of factors ($D=R$), while each of the loading functions is assumed to be linked to only one of the observed characteristics, i.e., $g_{\tau,r}(\bm{x}_i) = g_{\tau,r}(x_{ir})$ for $r=1,\ldots, R$. Note that these assumptions facilitate the interpretation of the estimated factors, e.g. the first estimated factor would be the value factor, the second one would be the momentum factor, and so on. However, both conditions also could be restrictive in other setups. For example, if $y_{it}$ represents the profit flow of firm $i$ at time $t$ and there are two factors capturing, say, a monetary shock and a fiscal shock, then it seems more reasonable to allow for dependence of the response of the firm's profit to the macro shocks on a wide range of firm characteristics --- such as size, leverage, growth, etc. --- that exceeds the number of factors. Moreover, a drawback of the two above-mentioned approaches is that they are more difficult to estimate and therefore require algorithms involving multiple iterations, particularly when the number of characteristics is large (say there are tens of characteristics, then there will be tens of factors). In contrast, in our setup, it is potentially easier to generalize CQFM to allow for high dimensional characteristics since the number of factors can be much smaller than the number of characteristics. Lastly, it should be noted that in the previous estimation methods, one needs to assume that the number of factors $R$ is a priori known, while in this paper we will propose a method that consistently estimates $R$ from the data at each quantile (see Section 3.3 below).  

Relative to the semiparametric factor models considered by \cite{fan2016projected}, the most salient difference is that the idiosyncratic errors in CQFM are subject to conditional quantile restrictions, rather than to conditional mean restrictions. From this perspective, as pointed out in \cite{chen2021quantile}, the QFM framework allows to recover different factor structures (including the factors, the loadings, and the number of factors) across different quantiles, even when the distribution of the idiosyncratic errors exhibits heavy tails. Hence, these features make CQFM a useful tool to analyze the co-movement of the financial market variables, where the correlation of the tail risks between different assets becomes the main object of interest. Furthermore, a relevant extension of \cite{fan2016projected} with respect to \cite{connor2012efficient} is that the factor loadings are allowed to be functions of other unobserved random variables, besides the set of observed characteristics. However, allowing for this more general case in the context of QFM would be very challenging. To see this, assume that 
\[\bm{\lambda}_i(\tau)  = \bm{g}_{\tau}(\bm{x}_i) +\bm{\gamma}_{i}, \] 
where $\bm{\gamma}_{i}$ is unobserved and independent of $\bm{x}_i$. Then model \eqref{eq1} can be written as 
\[ y_{it} = \bm{g}_{\tau}(\bm{x}_i)' \bm{f}_t(\tau)+\tilde{u}_{it}(\tau) \quad \text{ where } \quad \tilde{u}_{it}(\tau) =u_{it}(\tau)+\bm{\gamma}_{i}'\bm{f}_t(\tau).\]
The above-mentioned model can be viewed as a CQFM with \textit{measurement errors}, where the new error terms $\tilde{u}_{it}(\tau)$ no longer satisfy the conditional quantile restrictions, even when $\bm{x}_i$ and $\bm{\gamma}_{i}$ are independent. The insight is that imposing quantile conditional restrictions for the two elements of  $\tilde{u}_{it}(\tau)$ does not imply that this conditional restriction should hold for their sum. By contrast, if conditional expectation were applied, as in AFM, both error terms would have zero means allowing the application of PPCA. In fact, dealing with measurement errors is far from being a trivial issue even in standard quantile regressions (see e.g. \cite*{hausman2021errors}).
Thus, in what follows, the analysis will be  restricted to the case where the factor loadings are fully explained by the observed characteristics, as in \cite{ma2021estimation}, but allowing each factor loading to depend on a host of observable characteristics instead of just a single one. Several data generating processes (DGPs) discussed in the Monte Carlo simulations (reported in the online appendix) provide examples of CQFM models to be estimated by QPPCA.

\subsection{Estimators}
To simplify the notations even further, in the rest of the paper we suppress the $\tau$-subscripts in the model and use $\bm{g}(),\bm{G}(\cdot), \bm{F}, \bm{U}$ instead of $\bm{g}_{\tau}(\cdot),\bm{G}_{\tau}(\cdot), \bm{F}_\tau, \bm{U}_{\tau}$.

Write $\theta_{0t}(\bm{x}_i) =\bm{g}(\bm{x}_i)'\bm{f}_t= \sum_{r=1}^{R}  g_{r}(\bm{x}_{i}) f_{tr}=\sum_{r=1}^{R}  (\sum_{d=1}^{D}g_{rd}(x_{id}))f_{tr}$. Let $\Theta$ be a space of continuous functions such that $\theta_{0t} \in\Theta$ for all $t=1,\ldots,T$, while $\{\Theta_n\}$ is a sequence of sieve spaces approximating $\Theta$. In particular, let us consider the following finite dimensional linear spaces:
\[ \Theta_n = \left\{ h: \mathcal{X} \mapsto \mathbb{R}, \quad h(\bm{x})= \sum_{d=1}^{D}\sum_{j=1}^{k_n} a_{jd}\phi_j(x_{d}): (a_{11},\ldots, a_{jd},\ldots,a_{k_nD}) \in \mathbb{R}^{Dk_n}\right\}, \]
where $\mathcal{X}\subset \mathbb{R}^D$ is the support of $\bm{x}_i$, and $\phi_1,\ldots,\phi_{k_n} $ is a set of continuous basis functions. 
Write 
\[  \underbrace{ \bm{\phi}_{k_n}(\bm{x}_{i})}_{D k_n \times 1}=\left[\phi_{1}(x_{i1}),\ldots,\phi_{k_n}(x_{i1}), \ldots  \phi_{1}(x_{id}),\ldots,\phi_{k_n}(x_{id}),\ldots,\phi_{1}(x_{iD}),\ldots,\phi_{k_n}(x_{iD}) \right] '.\]
Suppose that for $r=1,\ldots,R$, there exists $\bm{b}_{01},\ldots,\bm{b}_{0R} \in \mathbb{R}^{Dk_n} $ such that for some constant $\alpha> 0$, 
\begin{equation}\label{approximation}
        \max_{1\leq r\leq R}\sup_{\bm{x} \in \mathcal{X}} \left| g_r(\bm{x}) -\bm{b}_{0r}' \bm{\phi}_{k_n}(\bm{x})\right| =O( k_n^{-\alpha})  . \end{equation}
Then, for $\bm{B}_0 =( \bm{b}_{01},\ldots,\bm{b}_{0R})\in\mathbb{R}^{Dk_n\times R}$, $\bm{a}_{0t} = \bm{B}_0 \bm{f}_t$ and $\pi_n \theta_{0t} (\cdot)=\bm{a}_{0t}'\bm{\phi}_{k_n}(\cdot)$, we have $\pi_n \theta_{0t} \in \Theta_n$ for all $t$ and 
\begin{equation}
\max_{1\leq t\leq T}\sup_{\bm{x} \in \mathcal{X}} \left| \pi_n \theta_{0t}(\bm{x})-\theta_{0t}(\bm{x})  \right| =O( k_n^{-\alpha}).
\end{equation}
Once the definitions above have been established, the next stage is to introduce our QPPCA estimation method which consists of the following three steps.

\vspace{0.3cm}

\noindent{\textbf{Step 1:}} Obtain the sieve estimator of $\theta_{0t}$. Let $\rho_{\tau}(u) =(\tau-\mathbf{1}\{u\leq 0\})u$ be the check function, and define $l(\theta, y_{it},\bm{x}_i) = \rho_{\tau}(y_{it} -\theta(\bm{x}_i))-\rho_{\tau}(y_{it} -\theta_{0t}(\bm{x}_i))$, $L_n(\theta) =n^{-1}\sum_{i=1}^{n}l(\theta, y_{it},\bm{x}_i) $. Then the sieve estimator $\hat{\theta}_{nt}$ is defined by
\[    L_n(  \hat{\theta}_{nt}) \leq \inf_{\theta\in\Theta_n} L_n( \theta).\]
In practice, $\hat{\theta}_{nt}$ can be obtained by a simple parametric quantile regression as follows: 
\begin{equation*}
 \hat{\bm{a}}_t  = \argmin_{ \bm{a} \in \mathbb{R}^{Dk_n}} \sum_{i=1}^{N} \rho_{\tau} \left( y_{it} - \bm{a}'\bm{\phi}_{k_n}(\bm{x}_i) \right) \quad \text{and} \quad \hat{\theta}_{nt}(\cdot) = \hat{\bm{a}}_t'\bm{\phi}_{k_n}(\cdot).
\end{equation*}

\vspace{0.3cm}

\noindent{\textbf{Step 2:}} Write $\hat{y}_{it} = \hat{\theta}_{nt}(\bm{x}_i)=\hat{\bm{a}}_t'\bm{\phi}_{k_n}(\bm{x}_i)$ and let $\hat{\bm{Y}}$ be the $n\times T$ matrix of $\hat{y}_{it}$. Then, the estimator of $\bm{F}$, denoted as $\hat{\bm{F}}$, is the matrix of eigenvectors (multiplied by $\sqrt{T}$) associated with $R$ largest eigenvalues of the $T\times T$ matrix $\hat{\bm{Y}}'\hat{\bm{Y}}$. Moreover, the estimator of the characteristics-based loading matrix $\bm{G}(\bm{X}) $ is given by $\hat{\bm{G}}(\bm{X}) = \hat{\bm{Y}} \hat{\bm{F}}/T$. It is well known that these estimators are the ones that minimize the objective function: $L_{nT}(\bm{G}(\bm{X}),\bm{F} ) = \| \hat{\bm{Y}} -  \bm{G}(\bm{X})'\bm{F} \|^2 $, subject to the standard normalizations, namely, $\bm{F}'\bm{F}/T=\bm{I}_R$ and $\bm{G}(\bm{X})'\bm{G}(\bm{X})/n$ is diagonal (see \cite{stock2002forecasting}).\footnote{ Note, however, that the estimator is invariant to the rotation transformations of the sieve bases.} 

\vspace{0.3cm}

\noindent{\textbf{Step 3:}} Estimate the factor loading functions: $g_r(\cdot)$ for $r=1,\ldots,R$. Let $\bm{A}_0 =(\bm{a}_{01},\ldots,\bm{a}_{0T})$ and $\hat{\bm{A}} =(\hat{\bm{a}}_{01},\ldots,\hat{\bm{a}}_{0T}) $. Intuitively, $\hat{\bm{A}}  \approx \bm{A}_0=\bm{B}_0\bm{F}' \approx  \bm{B}_0 \hat{\bm{F}}'$, as a result of which $\bm{B}$ can be simply estimated as 
\begin{equation}
\hat{\bm{B}} =  \hat{\bm{A}}\hat{\bm{F}}/T  .
\end{equation}
The estimator of $\bm{g}(\bm{x})$ for any $\bm{x}\in\mathcal{X}$ is given by $\hat{\bm{g}}(\bm{x})' = \bm{\phi}_{k_n}(\bm{x})' \hat{\bm{B}}$.

\vspace{0.1cm}

\begin{rem}
The basic idea of the projections is to smooth the observations $\{y_{it}\}_{1\leq i \leq n}$ for each given $t$ against its associated covariates, as in step 1, to then compute the factors of the projections in step 2; finally using the estimated factors and the estimates of the sieve functions in step 1, the loading functions are easily obtained in step 3. The main difference between this three-stage estimation method and the PPCA estimating approach of \cite{fan2016projected} is how we project $\bm{y}_t$ onto the space of $\bm{X}$ in the first step, namely, how $\bm{a}_{01},\ldots,\bm{a}_{0T}$ are estimated. The use of sieve quantile regressions instead of the least squares projections is a natural choice given that the idiosyncratic errors in CQFM are subject to conditional quantile restrictions. When the distributions of the errors are symmetric around 0, the QPPCA estimators at $\tau=0.5$ can be viewed as a robust version of the PPCA estimators since the consistency of the QPPCA estimators does not rely on moment restrictions of the errors (see Theorem 1 below). 
\end{rem}

\vspace{0.1cm}

\begin{rem}
The SQFA estimation method advocated by \cite{ma2021estimation} chooses $\bm{B}$ and $\bm{F}$ in an iterative fashion to minimize the following objective function:
\[L_{nT}(\bm{B},\bm{F} ) = \sum_{i=1}^{n}\sum_{t=1}^{T}\rho_{\tau} ( y_{it} - \bm{\phi}_{k_n}(\bm{x}_i)'\bm{B}\bm{f}_t ), \]
while the quantile factor analysis (\textbf{QFA}) proposed by \cite{chen2021quantile} for QFM relies on a similar approach which estimates the factor loadings $\bm{G}(\bm{X})$ and $\bm{F}$ jointly. Accordingly, both SQFA and QFA require $n$ and $T$ go to infinity in order to establish the consistency of the estimators. By contrast, as will be shown in the next section, the consistency of the QPPCA estimators can be established either when $T$ is fixed or $T$ goes to infinity along with $n$.
\end{rem}


\section{Asymptotic Properties of the Estimators}
In this section, we derive the rates of convergence and the asymptotic distributions of the QPPCA estimators proposed in the previous section. To simplify the discussion, the number of factors is taken to be known in the first two subsections while this assumption is relaxed in the last subsection, where a consistent estimator of $R$ is introduced.   

As in \cite{chen2021quantile} and \cite{ma2021estimation},  the quantile factors are treated as non-random constants in the asymptotic analysis. Hence, the conditional quantile restrictions on the idiosyncratic errors imply that 
\begin{equation}\label{quantile_res} 
P[ u_{it} \leq 0 |\bm{x}_i = \bm{x}]=\tau \text{ for any }\bm{x}\in\mathcal{X}. 
\end{equation}
Alternatively, all the assumptions and results to be presented below could be understood as being conditional on the realizations of the factors. 

Lastly, it should be noted that all the results to be presented below hold either when: (i) $T$ is fixed and $n\rightarrow\infty$, or (ii) $n,T\rightarrow\infty$. The first case is also called the \textit{high-dimension-low-sample-size} setup in the statistics literature (see \cite*{shen2013consistency} and \cite*{jung2009pca}). One of the main insights of \cite{fan2016projected} is that dimensionality is a blessing rather than a curse in the context of CFM, implying that their PPCA estimators are consistent even when $T$ is fixed. Our results below extend the finite-$T$-consistency results of \cite{fan2016projected} to CQFM.

\subsection{Rates of convergence}
Suppose that the observed data $\{y_{it}\}$ are generated by \eqref{eq3} and that $\{u_{it}\}$ satisfy \eqref{quantile_res}. Let 
\[ \varepsilon_n = \sqrt{k_n/n} \vee k_n^{-\alpha} \quad \text{ and }\quad \varepsilon_{nT} = \sqrt{\ln T}\vee 1 \cdot  \varepsilon_n.\]
For any $\theta_1,\theta_2\in\Theta$, define the pseudo-metric $d(\theta_1,\theta_2) \equiv \sqrt{\mathbb{E}\left( \theta_1(\bm{x}_i) -\theta_2(\bm{x}_i)\right)^2 }$. 
The following set of conditions are required to establish the uniform rate of convergence of $\hat{\theta}_{n1},\ldots,\hat{\theta}_{nT}$, which is a crucial result to prove the other theorems.

\begin{ass}Let $M$ be a generic bounded constant.\\ 
(i) Define $\bm{z}_i =(u_{i1},\ldots,u_{iT},\bm{x}_i)$. Then, $\bm{z}_1,\ldots,\bm{z}_n$ are i.i.d. Moreover, the distributions of $(u_{i1},\bm{x}_i),\ldots, (u_{iT},\bm{x}_i)$ are identical for each $i$.   \\
(ii) Equation \eqref{approximation} holds for some $\alpha\geq1$.\\
(iii) $\mathcal{X}\subset \mathbb{R}^D$ is bounded, and $\sup_{\theta\in\Theta}\sup_{\bm{x}\in\mathcal{X}}|\theta(\bm{x})|<M$. $\|\bm{f}_t\|<M$ for all $t=1,\ldots,T$. \\
(iv) The conditional density of $u_{it}$ given $\bm{x}_i=\bm{x}$, denoted as $\mathsf{f}(\cdot|\bm{x})$, satisfies: $ 0< \inf_{\mathcal{X}} \mathsf{f}(0|\bm{x})
\leq \sup_{\mathcal{X}} \mathsf{f}(0|\bm{x})<\infty$ and $\sup_{\mathcal{X}} |\mathsf{f}(z|\bm{x})-\mathsf{f}(0|\bm{x}) | \rightarrow 0 $ as $|z|\rightarrow 0$.  \\
(v) As $n\rightarrow\infty$, $k_n\rightarrow \infty$ and $ \varepsilon_{nT} \rightarrow 0$.
\end{ass}

Although, in principle, Assumption 1(i) is stronger than those in \cite{fan2016projected} and \cite{ma2021estimation}, it can be relaxed to allow for weak cross-sectional dependence --- see Remark 3 below for the details. Assumption 1(ii) is a general condition on the sieve approximations that can be easily verified using more primitive conditions. For instance, it holds if $\Theta$ is an $\alpha-$smooth H\"{o}lder space (see \cite{chen2007large} for further examples). Assumption 1(iii) and Assumption 1(iv) are also standard in sieve quantile regressions, noting that the last assumption imposes very mild restrictions on the size of $T$ when it goes to infinity jointly with $n$. 

\begin{prop}
Under Assumption 1, when either $T$ is fixed or $T\rightarrow\infty$ as $n\rightarrow\infty$, it holds that  $\max_{1\leq t\leq T}d(\hat{\theta}_{nt},\theta_{0t})=O_P(\varepsilon_{nT})$. 
\end{prop}

\vspace{0.1cm}
\begin{rem}
The proof of Proposition 1 is based on Corollary 1 of \cite{chen1998sieve}. In particular, we show that 
\[ P \left[ \max_t d(\hat{\theta}_{nt},\theta_{0t}) \geq C \varepsilon_{nT}\right] 
\leq \sum_{t=1}^{T} P \left[  d(\hat{\theta}_{nt},\theta_{0t}) \geq C \varepsilon_{nT}\right]  
\leq c_1 \exp\left\{ C^2  \ln T (1-c_2 n\varepsilon_n^2) \right\}\]
for any $C\geq 1$ and some constants $c_1,c_2$. Moreover, as shown in \cite{chen1998sieve}, the above inequality holds when the observations are generated from a stationary uniform ($\phi$-) mixing sequence with $\phi(j)\lesssim j^{-\zeta}$ for some $\zeta>1$. Thus, in line with \cite{connor1993test}, \cite{lee2016series} and \cite{ma2021estimation}, one can assume the existence of a reordering of the cross-sectional units such that their dependence can be characterized by the uniform mixing condition mentioned above, and the conclusion of Proposition 1 will still hold.
\end{rem}

To establish the convergence rates of the estimated factors and loading functions, some further assumptions are required.

\begin{ass} Let $M$ be a generic bounded constant. \\
(i) Let $\bm{\Sigma}_{\phi} =\mathbb{E}[ \bm{\phi}_{k_n}(\bm{x}_i)\bm{\phi}_{k_n}(\bm{x}_i)']$. Then, there exist constants $c_1,c_2$ such that $0 < c_1\leq \lambda_{\min}\left(  \bm{\Sigma}_{\phi} \right)\leq \lambda_{\max}\left(  \bm{\Sigma}_{\phi} \right) \leq c_2<\infty$ for all $n$.\\
(ii) $k_n^2/n\rightarrow 0$ as $n\rightarrow\infty$.\\
(iii) There exist a constant $c>0$ such that $\lambda_{\min}( \bm{F}'\bm{F}/T)>c$ for all $T$.\\
(iv) $\hat{\bm{\Sigma}}_g \equiv n^{-1}\sum_{i=1}^{n}\bm{g}(\bm{x}_i)\bm{g}(\bm{x}_i)' \overset{P}{\rightarrow} \bm{\Sigma}_g>0$ as $n\rightarrow\infty$.\\
(v) The eigenvalues of $\bm{\Sigma}_g \cdot \bm{F}'\bm{F}/T $ are distinct.
\end{ass}

The conditions in Assumption 2 are all standard in the literature on factor models and sieve estimation. In particular, Assumption 2(ii) strengthens Assumption 1(v), and Assumption 2(iii) implicitly requires that $T\geq R$. In comparison, Assumption A0 of \cite{ma2021estimation} imposes that $\lim \inf _{T\rightarrow \infty}|T^{-1}\sum_{t=1}^{T}f_{tr} |>0$ for all $r=1,\ldots, R$, which excludes the possibility that the underlying time series generating $\bm{F}$ has zero mean. The following theorem gives the rates of convergence of the estimated factors and loading functions.

\begin{thm}
Let $\hat{\bm{\Omega}}$ be the diagonal matrix whose elements are the eigenvalues of $\hat{\bm{Y}} '\hat{\bm{Y}}/(nT)$, and define $\hat{\bm{H}}=\hat{\bm{\Sigma}}_g ( \bm{F}' \hat{\bm{F}}/T)\hat{\bm{\Omega}}^{-1}$. Then, under Assumptions 1 and 2, the following results hold either when $T$ is fixed or $T\rightarrow\infty$ as $n\rightarrow\infty$,: \\
(i) $\|\hat{\bm{F}}-\bm{F}\hat{\bm{H}}\|/\sqrt{T}=O_P(\varepsilon_{nT})$.\\
(ii) $\|\hat{\bm{G}}(\bm{X}) - \bm{G}(\bm{X})( \hat{\bm{H}}')^{-1}\|/\sqrt{n}=O_P(\varepsilon_{nT})$.\\
(iii) $\sup_{\bm{x}\in \mathcal{X}}\| \hat{\bm{g}}(\bm{x}) -\hat{\bm{H}}^{-1} \bm{g}(\bm{x}) \|   = O_P(  \sqrt{k_n}\varepsilon_{nT})$.
\end{thm}

A few remarks on this result are relevant. First, it is worth highlighting that Theorem 1 (and Theorem 2 below) is obtained without imposing any restrictions on the time-series dependence of the idiosyncratic errors, while both \cite{fan2016projected} and \cite{ma2021estimation} impose some kind of weak-time-series-dependence conditions. Second, while our setup does not require any moment restrictions on $u_{it}$, Assumption 3.4 of \cite{fan2016projected} needs the error terms to have exponential tails. Third, the price to pay for these nice properties is that the convergence rates given in Theorem 1 are generally slower than those of \cite{fan2016projected}, mainly due to our use of sieve quantile estimators rather than sieve least square estimators. In fact, in the proof of Theorem 1, we only use the uniform convergence rate of $\hat{\bm{a}}_t$ (see Lemma 1 in the online appendix). Yet, by exploring the Bahadur representation of $\hat{\bm{a}}_t$, the convergence rate of the estimated loading functions can be improved when $T$ is large (see Theorem 3 below), whilst the convergence rate of the estimated factors can be greatly improved even when $T$ is fixed if the following extra assumptions are imposed. 

\begin{ass} Let $L$ be a generic bounded constant and let $\mathsf{f}(\cdot)$ denote the p.d.f. of $u_{it}$. \\
(i) For each $i$, $\bm{x}_i$ is independent of $(u_{i1},\ldots,u_{iT})$. \\
(ii) $ \left| \mathsf{f}(c)-\mathsf{f}(0) \right| \leq L | c|$ for any $c$ in a neighborhood of $0$.\\
(iii) Equation \eqref{approximation} holds for some $\alpha\geq 3$.
\end{ass}

Assumption 3(i) essentially requires that the observed characteristics only affect the location but not the scale of the distributions of $y_{it}$. In such a case, the leading term in the Bahadur representation of $\hat{\bm{a}}_t$ has a similar structure to the least square estimators (see Lemma 2 in the online appendix). Thus, this assumption implies an improved convergence rate of $\hat{\bm{F}}$ which happens to be as fast as that of the PPCA estimators (see Theorem 4.1 of \cite{fan2016projected}).

\begin{thm} Let $\eta_{nT}=\sqrt{\ln (  k_n^{-1/4}\varepsilon_{nT}^{-1/2})}\cdot k_n^{5/4}\varepsilon_{nT}^{1/2} n^{-1/2} $. Under Assumptions 1 to 3, we have
\[ \| \hat{\bm{F}} - \bm{F}\hat{\bm{H}}\|/\sqrt{ T}  = O_P\left(n^{-1/2}\vee k_n^{-\alpha} \vee  \eta_{nT} \vee \varepsilon_{nT}^2\right). \]
Moreover, if $T\asymp n^{\gamma_1}$ and $ k_n \asymp n^{1/(6+\gamma_2)}$ for some $\gamma_1\geq 0$ and $\gamma_2>0$, then 
\[\| \hat{\bm{F}} - \bm{F}\hat{\bm{H}}\|/\sqrt{T}  = O_P\left(n^{-1/2}\vee k_n^{-\alpha} \right).\] 
\end{thm}

\vspace{0.1cm}
\begin{rem}
The term $\eta_{nT}$ in Theorem 2 represents the higher-order terms in the Bahadur representation of  $\hat{\bm{a}}_t$. When $\alpha$ is large, $\eta_{nT}$ is approximately equal to $k_n^{3/2}n^{-3/4}$. Note that this slightly unusual expression of $\eta_{nT}$ is mainly due to the non-smoothness of the check function. Indeed, similar terms can be found in Theorem 2 of \cite{horowitz2005nonparametric}, Theorem 3.2 of \cite{kato2012asymptotics} and Theorem 2 of \cite{ma2021estimation}.
\end{rem}

\subsection{Asymptotic distribution}
Define $\bm{\Sigma}_{\mathsf{f}\phi}=\mathbb{E}[\mathsf{f}(0|\bm{x}_i) \bm{\phi}_{k_n}(\bm{x}_i)\bm{\phi}_{k_n}(\bm{x}_i)' ]$ and $\sigma_{k_n}^2=\bm{\phi}'_{k_n}(\bm{x})\bm{\Sigma}_{\mathsf{f}\phi}^{-1} \bm{\Sigma}_{\phi} \bm{\Sigma}_{\mathsf{f}\phi}^{-1}\bm{\phi}_{k_n}(\bm{x}).$

\begin{ass} Let $L$ be a generic bounded constant. \\
(i) $u_{i1},\ldots,u_{iT}$ are independent conditional on $\bm{x}_i$.\\
(ii) $ \left| \mathsf{f}(c|\bm{x})-\mathsf{f}(0|\bm{x}) \right| \leq L | c|$ for any $c$ in a neighborhood of $0$ and any $\bm{x}\in\mathcal{X}$.\\
(iii) There exist constants $c_1,c_2$ such that $0<c_1 \leq  \lambda_{\min}( \bm{\Sigma}_{\mathsf{f}\phi}) \leq \lambda_{\max}( \bm{\Sigma}_{\mathsf{f}\phi})\leq   c_2<\infty $ for all $k_n$. \\
(iv) $(nT)^{1/2}k_n^{1/2-\alpha}\sigma_{k_n}^{-1}=o(1)$ and $(nT)^{1/2} k_n^{1/2} \eta_{nT}\sigma_{k_n}^{-1}=o(1)$.
\end{ass}

Assumption 4(i) is adopted  for simplicity, though it could be replaced by $\beta-$mixing conditions at the cost of getting more complex asymptotic covariance matrices. When $\sigma_{k_n} \asymp k_n^{1/2}$ and $T$ is fixed, Assumption 4(iv) essentially requires that $n^{1/2}k_n^{-\alpha}=o(1)$ and $n^{1/2}\eta_{nT}=o(1)$, or $k_n^6\ll n \ll k_n^{2\alpha}$. As a result, we need \eqref{approximation} to hold with $\alpha>3$. The remaining conditions in Assumption 4 are standard --- see, e.g. Assumptions 3 and 5 of \cite{horowitz2005nonparametric}. We are now in the position of establishing the asymptotic distribution of the estimated loading functions.

\vspace{0.1cm}

\begin{thm} Under Assumptions 1, 2, and 4, it holds that for any $\bm{x}\in\mathcal{X}$ 
\[\bm{\Sigma}_{T,\tau}^{-1/2}(\hat{\bm{H}}')^{-1} \cdot  \frac{\sqrt{nT}}{\sigma_{k_n}}\left( \hat{\bm{g}}(\bm{x}) - ( \bm{F}' \hat{\bm{F}}/T)'\bm{g}(\bm{x}) \right) \overset{d}{\rightarrow} N(0, \bm{I}_R ), \]
where $  \bm{\Sigma}_{T,\tau} =\tau(1-\tau) (\bm{F}'\bm{F}/T)$.
\end{thm}

The asymptotic distribution of $\hat{\bm{F}}$ is more difficult to derive, especially when $n$ and $T$ go to infinity simultaneously. For this reason, instead of focusing on $\hat{\bm{F}}$, let us consider the following updated estimator for the factors:
\[ \tilde{\bm{F}} =  \hat{\bm{Y}}' \hat{\bm{G}}(\bm{X}) \cdot ( \hat{\bm{G}}(\bm{X})' \hat{\bm{G}}(\bm{X}) )^{-1}.\]
In addition, let $\tilde{\bm{H}}= ( \bm{G}(\bm{X})'\hat{\bm{G}}(\bm{X}) /n) ( \hat{\bm{G}}(\bm{X})' \hat{\bm{G}}(\bm{X})/n )^{-1}$ and
\[ \bm{\Xi}_{\tau}=\tau(1-\tau)\cdot\bm{\Sigma}_{g}^{-1}\mathbb{E}[\bm{g}(\bm{x}_i) \bm{\phi}_{k_n}(\bm{x}_i)']\bm{\Sigma}_{\mathsf{f}\phi}^{-1}
\bm{\Sigma}_{\phi}\bm{\Sigma}_{\mathsf{f}\phi}^{-1}\mathbb{E}[ \bm{\phi}_{k_n}(\bm{x}_i)\bm{g}(\bm{x}_i)']\bm{\Sigma}_{g}^{-1}.
\]
Then, the following additional assumption is needed to derive the asymptotic distribution of $\tilde{\bm{f}}_t $.
\begin{ass} Conditions (i) to (iii) of Assumption 4 hold and $n^{1/2}k_n^{-\alpha}/\| \bm{\Xi}_{\tau}\|^{1/2}=o(1)$, $n^{1/2}\eta_{nT} /\|\bm{\Xi}_{\tau}\|^{1/2}=o(1)$, $\varepsilon_{nT}\sqrt{k_n}=o(1)$.
\end{ass}

\begin{thm} Under Assumptions 1, 2, and 5, it holds for all $t=1,\ldots,T$, 
\[ \bm{\Xi}_{\tau}^{-1/2 } (\hat{\bm{H}}')^{-1} \sqrt{n}( \tilde{\bm{f}}_t - \tilde{\bm{H}}'\bm{f}_t)  \overset{d}{\rightarrow}N(0,\bm{I}_R  ).\] 
\end{thm}

When $\|\bm{\Xi}_{\tau}\|\asymp k_n$, the convergence rate of $\tilde{\bm{f}}_t$ is $O_P(\sqrt{n/k_n})$, and Assumption 5 requires that $n^{1/2}k_n^{-\alpha-1/2}=o(1)$ and $n^{1/2}\eta_{nT}k_n^{-1/2}=o(1)$, or $k_n^4\ll n \ll k_n^{2\alpha+1}$. As a result, we need \eqref{approximation} to hold with $\alpha\geq 2$. Alternatively, if Assumption 3(i) holds, it can be shown that
\[  \| \bm{\Xi}_{\tau} -\tau(1-\tau)\cdot\bm{\Sigma}_{g}^{-1} \cdot \mathsf{f}^{-2}(0) \| =O(k_n^{-\alpha}).\]
In this case, the convergence rate of $\tilde{\bm{f}}_t$ is $\sqrt{n}$ for each $t$.

\vspace{0.1cm}

\begin{rem} As in Proposition 1 of \cite{bai2003inferential}, it can be shown that $\hat{\bm{H}}$, $\tilde{\bm{H}}$ and $ \bm{F}' \hat{\bm{F}}/T$ all converge in probability to some positive definite matrices as $n,T\rightarrow\infty$. In particular, if $\bm{F}' \bm{F}/T =\bm{I}_R$ and $\hat{\bm{\Sigma}}_g$ is diagonal, the probability limits of $\hat{\bm{H}}$, $\tilde{\bm{H}}$ and $\bm{F}' \hat{\bm{F}}/T$ are all equal to $\bm{I}_R$.
\end{rem}

\vspace{0.1cm}

\begin{rem} Note that both Theorems 3 and 4 are fulfilled when $T$ is fixed as well as when $T\rightarrow\infty$ as $n\rightarrow\infty$. In the latter case, if $n\asymp T$, \cite{chen2021quantile} show that the estimators of the quantile factors are $\sqrt{n}$-consistent and asymptotically normal under more general conditions. Thus, whenever $T$ is as large as $n$ and the quantile factors are the main objects of interest, these estimators should be preferable. However, if $T$ is small and $n$ is large, Theorem 4 above shows that the QPPCA estimators proposed here remain consistent and asymptotically normal.
\end{rem}

\subsection{Estimating the number of factors}
Given that $\hat{\bm{Y}}=\bm{\Phi}(\bm{X})\bm{\hat{A}} \approx \bm{\Phi}(\bm{X})\bm{A}_0  =\bm{\Phi}(\bm{X}) \bm{B}_0 \bm{F}' \approx \bm{G}(\bm{X})  \bm{F}'$, the rank of $\hat{\bm{Y}}$ is asymptotically equal to $R$. Let $\hat{\rho}_1,\ldots,\hat{\rho}_{\bar{R}}$ be the $\bar{R}$ largest eigenvalues of $ \hat{\bm{Y}}\hat{\bm{Y}}'/(nT)$ in descending order. Then, the estimator of $R$ is given by the number of non-vanishing eigenvalues of $ \hat{\bm{Y}}\hat{\bm{Y}}'/(nT)$, i.e.
\begin{equation}\label{R_estimator}
 \hat{R} = \sum_{j=1}^{\bar{R}} \mathbf{1}\{ \hat{\rho}_j >p_n\},
\end{equation}
where $\{p_n\}$ is a sequence of non-increasing positive constants. The following theorem provides conditions on the threshold $p_n$ to establish the consistency of $\hat{R}$ which, following \cite{chen2021quantile}, is denoted as the rank minimization estimator of the number of factors.

\begin{thm}
Suppose that $\bar{R} \geq R$ and $p_n\rightarrow 0$, $p_n\varepsilon_{nT}^{-1}\rightarrow\infty$ as $n\rightarrow \infty$, then under Assumptions 1 and 2, we have $P[\hat{R}=R] \rightarrow 1$ as $n\rightarrow \infty$.
\end{thm}

To prove Theorem 5, we show that the largest $R$ eigenvalues of $ \hat{\bm{Y}}\hat{\bm{Y}}'/(nT)$ converge in probability to some positive constants, while the remaining eigenvalues are all $O_P( \varepsilon_{nT})$. Then, the decreasing sequence $\{p_n\}$ is chosen to dominate the vanishing eigenvalues in the limit. Again, this result also holds even when $T$ is fixed. 

In theory, the choice of $p_n$ is determined by $\alpha$, which depends on the smoothness of the unknown quantile loading functions. Thus, a conservative choice of $p_n$ can rely on assuming that $\alpha =  1$. In this case, $\varepsilon_{nT}   = ( k_n^{1/2}n^{-1/2} \vee k_n^{-1} )\ln T $, and the optimal choice of $k_n$ is $k_n^{\ast} \asymp n^{1/3}$. Hence, to satisfy the condition of Theorem 5, we need $p_n \gg n^{-1/3}\ln T$. The following choice is recommended in practice:
\begin{equation} \label{eq9}
p_n =d \cdot \hat{\rho}_1^{1/2} \cdot n^{-1/4} \ln T,
\end{equation}
where $d$ is a positive constant and $ \hat{\rho}_1^{1/2}$ plays the role of a normalization factor. 

\begin{rem}
Alternatively, to avoid the choice of the threshold sequence $\{p_n\}$, use could be made of the approach proposed by  \cite{ahn2013eigenvalue} to estimate the number of factor by maximizing the ratios of consecutive eigenvalues, i.e. 
\[    \tilde{R} = \argmax_{j=1,\ldots,\bar{R}} \frac{ \hat{\rho}_j }{ \hat{\rho}_{j+1}}.     \]
This is the estimator considered by \cite{fan2016projected} for AFM where the error terms are required to be sub-Gaussian. For this reason, a formal proof of the consistency of this estimator in the context of QFM is technically challenging and is left for further research.
\end{rem}
 
 \section{Simulations}
In this section, we summarize results from a set of Monte Carlo simulations to study the behavior in finite samples of the QPPCA estimators regarding the estimation of the number of factors, the factors themselves and their loading functions. We generate random draws of the characteristics $\{x_{id},d=1,\ldots,D\}$, considering several combinations of their number $D$ and the number of factors $R$. To save space, the main findings are summarized here, relegating the description of the DGPs (location-shift models with one or two factors shifting the mean and another factor shifting the variance), tables and figures to the online appendix.  

First, as regards the estimation of the number of factors, we find that: (i) the rank-minimization (Table A.1) and the eigen-ratio (Table A.2) criteria estimate accurately the number of factors for small $T$ and large $n$ samples, supporting the  previous claim about their consistency even when $T$ is fixed (Table A.1); (ii) both estimators perform well when $n$ is large (=1000) even when the errors follow the standard Cauchy distribution, providing this time support for the claim that the rank-minimization estimator is consistent even in the absence of moment restrictions on the error terms; and (iii) while both estimators yield similar results when $n=1000$, the rank minimization estimator outperforms the eigen-ratio estimator when $n$ is not sufficiently large. 

Second, we run simulations about the estimation of factors comparing four competing methods: (i) QQPCA; (ii) the quantile factor analysis estimator (QFA) of \cite{chen2021quantile}; (iii) the PPCA estimator proposed by \cite{fan2016projected}; and (iv) the standard PCA estimator of \cite{bai2002determining} for AFM. Regarding the choices of $n$ and $T$, two different scenarios are considered: one with fixed $T=10,50$ and $n$ increasing from 50 to 500, and another with fixed $n=100,200$ and $T$ increasing from 5 to 200. The main findings are as follows: (i) PPCA and QPPCA perform better when the error terms are normal or t(3), whereas QPPCA performs much better when it is Cauchy (Figures A.1 and A.2), a case where, not surprisingly, PCA fails; (ii) all methods estimate the location factors well, except when errors are heavy-tailed (Tables A.3 to A.5); (iii) QPPCA fares much better than SQFA when the number of characteristics exceeds the number of factors (Tables A.6 and A.7). 

Finally, regarding the estimation of the loading functions (Figures A.3 to A.6), QPPCA exhibits good performance even when $T$ is not large.

 \section{Empirical Application}

In this section, the QPPCA estimation method is applied to investigate the factor structure of security returns. Following \cite{fan2016projected}, we use a dataset that includes information on the daily returns of S\&P500 index securities with complete daily closing price records from 2005 to 2013.\footnote{This dataset is downloaded from CRSP (Center for Research in Security Prices).} The sample consists of 355 stocks, whose book value and market capitalization are drawn from Compustat. Moreover, as is conventional in this literature, the 1-month US treasury bond rate is used as the risk-free rate to compute the daily excess return of each stock.

Following \cite{connor2012efficient}, \cite{fan2016projected}, and \cite{ma2021estimation}, four characteristics are considered: \textit{size, value, momentum} and \textit{volatility}, which are standardized to have zero means and unit standard deviations. Similar to \cite{fan2016projected}, we analyze the data corresponding to the first quarter of 2006, which includes $T=62$ observations. The second Chebyshev polynomials are used as the basis functions in the sieve regressions and $k_n=4$. 
  	
First, Table 1 shows the estimated number of mean factors using the eigen-ratio estimator proposed by \cite{fan2016projected} and the estimated numbers of quantile factors using the QPPCA rank minimization estimator for $\tau\in\{0.05,0.25,0.5,0.75,0.95\}$\footnote{The threshold $p_n$ is chosen as in \eqref{eq9}, with $d=1/4$.}. The five largest eigenvalues of $\hat{Y}\hat{Y}'$ and the threshold $p_n$ are also displayed in this table. In addition, the estimated numbers of quantile factors using the QFA rank minimization estimator are reported in the last column. Overall, the results provide strong evidence in favor of the existence of a single location factor and one factor at each quantile.

Second, Table 2 shows the correlation coefficients between the estimated location factors by PPCA and the estimated quantile factors by QPPCA for the above-mentioned values of $\tau$. The sample means of each estimated factor are also reported in the last column. Figure 1 in turn provides plots of these factors which exhibit different means and high correlations. As can be inspected, the PPCA factor is highly correlated with the QPPCA factor corresponding to $\tau=0.5$, but misses the factors at the more extreme quantiles.

Third, Figure 2 shows the estimated loading functions of the four characteristics using PPCA and QPPCA at $\tau=0.5$ where values of each standardized covariate appears in the horizontal axis. The fact that both methods yield similar estimated loading functions indicates that the idiosyncratic errors of the stock returns have symmetric distributions so that the mean and the median coincide. 
	
Fourth, given its better performance, Figure 3 plots the estimated loading functions of the four characteristics under consideration using QPPCA at different quantiles. In general, these functions feature considerable variation both across the values of the characteristics and the quantiles. A few salient findings emerge. First, the loading functions of size and volatility  seem to behave monotonically at all quantiles while, for value and momentum, they exhibit a clear non-linear pattern, mostly looking U-shaped. By the way, the shapes of the loading function resemble those reported by \cite{ma2021estimation} except for value (i.e. a value stock refers to shares of a company that appears to trade at a lower price relative to its fundamentals, such as dividends, earnings, or sales) which these authors find to have an inverted U-shape. Next, for all characteristics, their quantile loading functions at $\tau=0.25$ and $\tau=0.5$ are very close. Lastly, there is strong evidence that the loading functions at the tails ($\tau=0.05,0.95$) have greater curvatures than  at the remaining quantiles. In sum, this empirical evidence points out that the estimated loading functions vary substantially across different quantiles, a fact that cannot be uncovered using the PPCA method. Yet, this is a useful finding since deviating from the efficient market hypothesis, factors contributing to alpha generation can have different relevance depending on the distribution of excess returns. Thus, the standard asset valuation techniques based on CAPM and the Fama-French factors should take these features into consideration to create a portfolio delivering excess returns over time beating the market. 

Finally, it is worth highlighting that QPPCA allows estimating the conditional quantile of excess returns $\mathsf{Q}_{\tau}[y_{it}|\bm{x}_{i}] = \bm{g}_{\tau}(\bm{x}_i)'\bm{f}_t$, yielding $\hat{y}_{it}(\tau) = \hat{\bm{a}}_t'\bm{\phi}_{k_n}(\bm{x}_i)$ as its estimator, where $\hat{\bm{a}}_t$ is obtained from the cross-sectional quantile regressions in step 1 of the three-step procedure introduced in Section 2. One could interpret $\hat{y}_{it}(\tau)$ as the ``quantile returns'' which is interesting from the perspective of empirical applications since it is idiosyncratic free, that is, much less noisy than the realized return $y_{it}$. Just as the literature on asset pricing has increasingly appreciated the concept of ``expected returns'' because it is noiseless (see e.g. \cite{elton1999presidential}), ``quantile returns'' are also interesting on their own and could perhaps help provide a better explanation of the distribution of returns, an issue which remains high in our research agenda.

\begin{table}[H]
			\centering
			\begin{threeparttable}
			\caption{Estimated numbers of factors}
			\label{Table:8}
			\begin{tabular}{cccccccccc}
				\toprule
				&       & \multicolumn{5}{c}{Five largest eigenvalues of $\hat{Y}\hat{Y}'$}  & $p_n$    & $\hat{r}$ & $\hat{r}_{QFA}$ \\ \hline
				mean(PPCA) &       & 0.929    & 0.090  & 0.081  & 0.066  & 0.043  &       & 1 &  \\ \hline
				quantile & $\tau$=0.5  & 0.887    & 0.094  & 0.084  & 0.053  & 0.043  & 0.224 & 1 & 1 \\ 
				& $\tau$=0.25 & 1.713    & 0.110  & 0.098  & 0.059  & 0.047  & 0.311 & 1 & 1 \\ 
				& $\tau$=0.75 & 2.706    & 0.115  & 0.087  & 0.074  & 0.067  & 0.391 & 1 & 1 \\ 
				& $\tau$=0.05  & 8.415    & 0.311  & 0.173  & 0.161  & 0.138  & 0.690 & 1 & 1 \\ 
				& $\tau$=0.95  & 13.715    & 0.567  & 0.428  & 0.291  & 0.246  & 0.880 & 1 & 1 \\ 
				\bottomrule
			\end{tabular}
\small
\begin{tablenotes}
\item Note: this table shows the estimated numbers of factors using the eigen-ratio estimator proposed by \cite{fan2016projected}, the proposed estimator in this paper, and the rank-minimization estimator proposed by \cite{chen2021quantile} for different $\tau$s. Column 3 to Column 7 give the 5 largest eigenvalues of $\hat{Y}\hat{Y}'$, where $\hat{Y}$ is the matrix of fitted values in the first-step sieve regressions, and $p_n$ is the threshold value defined in \eqref{eq9}.
\end{tablenotes}
\end{threeparttable}
\end{table}

\begin{table}[H]	
			\centering
			\begin{threeparttable}
			\caption{Correlations and means of estimated factors}
			\label{Table:9}
			\begin{tabular}{ccccccc|c}
				\toprule
				& $\tau$=0.05 & $\tau$=0.25 & $\tau$=0.5  & $\tau$=0.75 & $\tau$=0.95 & PPCA & Mean  \\
				$\tau$=0.05 & 1     & 0.922 & 0.852 & 0.767 & 0.611 & 0.863 & 0.943 \\
				$\tau$=0.25 &       & 1     & 0.975 & 0.924 & 0.753 & 0.973 & 0.738 \\
				$\tau$=0.5  &       &       & 1     & 0.971 & 0.814 & 0.990 &-0.121 \\
				$\tau$=0.75 &       &       &       & 1     & 0.877 & 0.979 &-0.784 \\
				$\tau$=0.95 &       &       &       &       & 1     & 0.862 &-0.943 \\
				PPCA  &       &       &       &       &       & 1 & -0.231     \\ 
				\bottomrule
			\end{tabular}
			\small
\begin{tablenotes}
\item Note: this table shows the correlations and sample means of the estimated mean factor using PPCA and the estimated quantile factors at different $\tau$s using QPPCA.
\end{tablenotes}
			\end{threeparttable}
		\end{table}
  
		\begin{figure}[H]
			\caption{Estimated factors using QPPCA for different quantiles}\label{figure:7}
			\centering\includegraphics[height=3.5in]{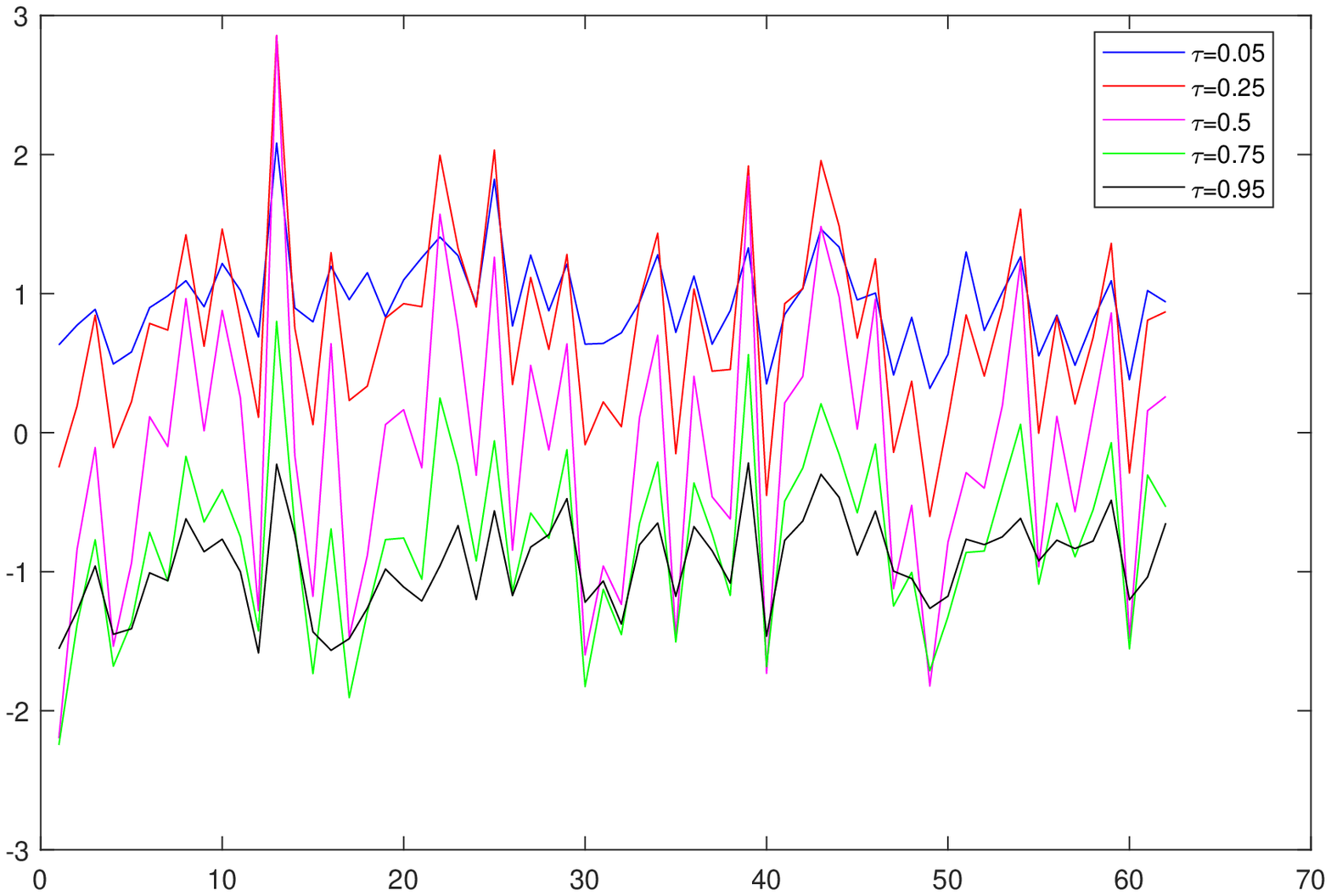}
				\captionsetup{singlelinecheck=off}
	\caption*{Note: This figure plots the estimated quantile factors at different quantiles using QPPCA.}
		\end{figure}
		
		\begin{figure}[H]
			\caption{Estimated loading functions using PPCA and QPPCA for $\tau=0.5$}\label{figure:8}
			\centering\includegraphics[width=7in]{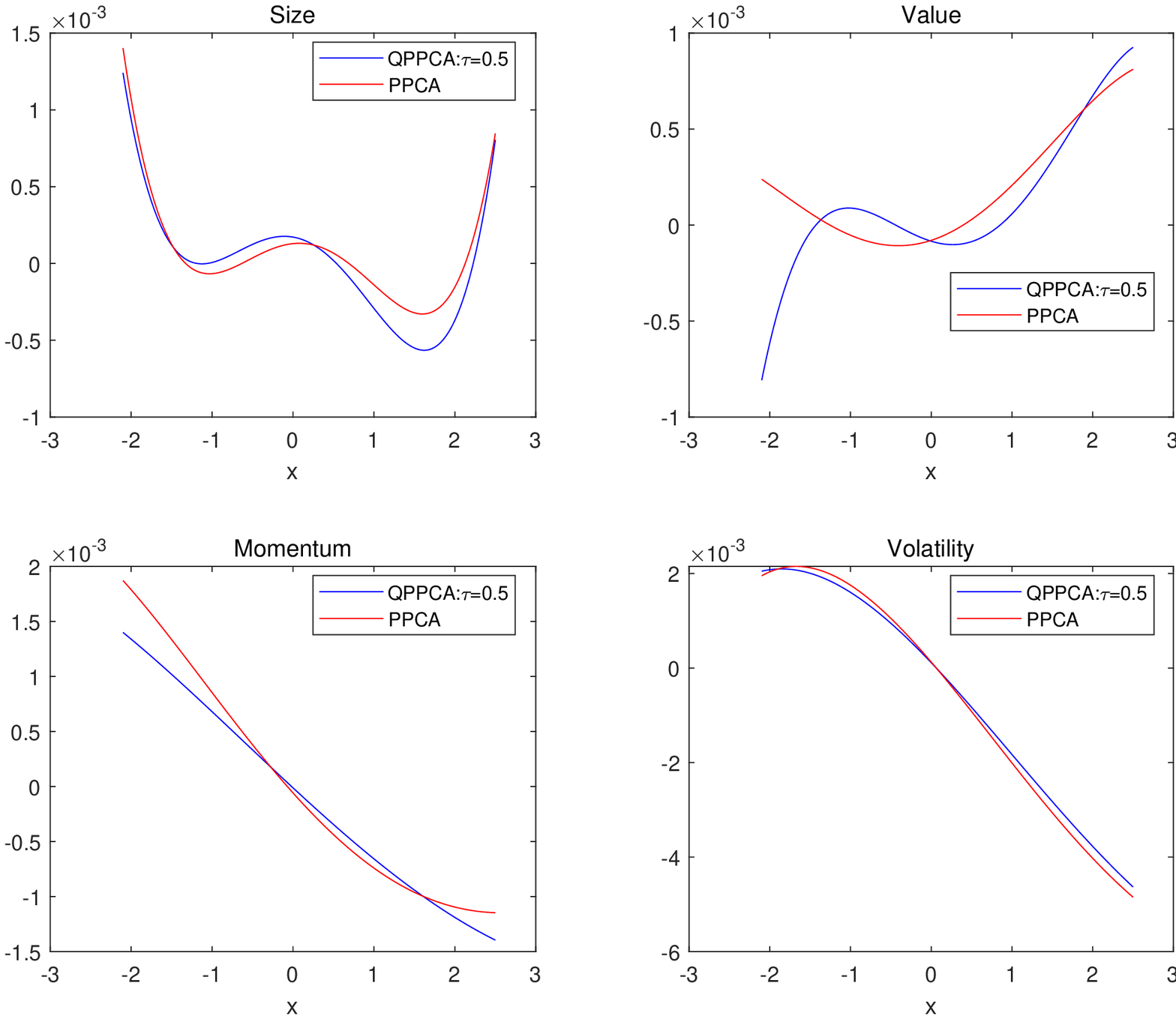}
			\captionsetup{singlelinecheck=off}
	\caption*{Note: this figure plots the estimated quantile factor loading functions of the four characteristics using PPCA and QPPCA at $\tau=0.5$}
			\end{figure}
			
		\begin{figure}[H]	
			\caption{Estimated loading functions using QPPCA for different quantiles} \label{figure:9}
			\centering\includegraphics[height=5in]{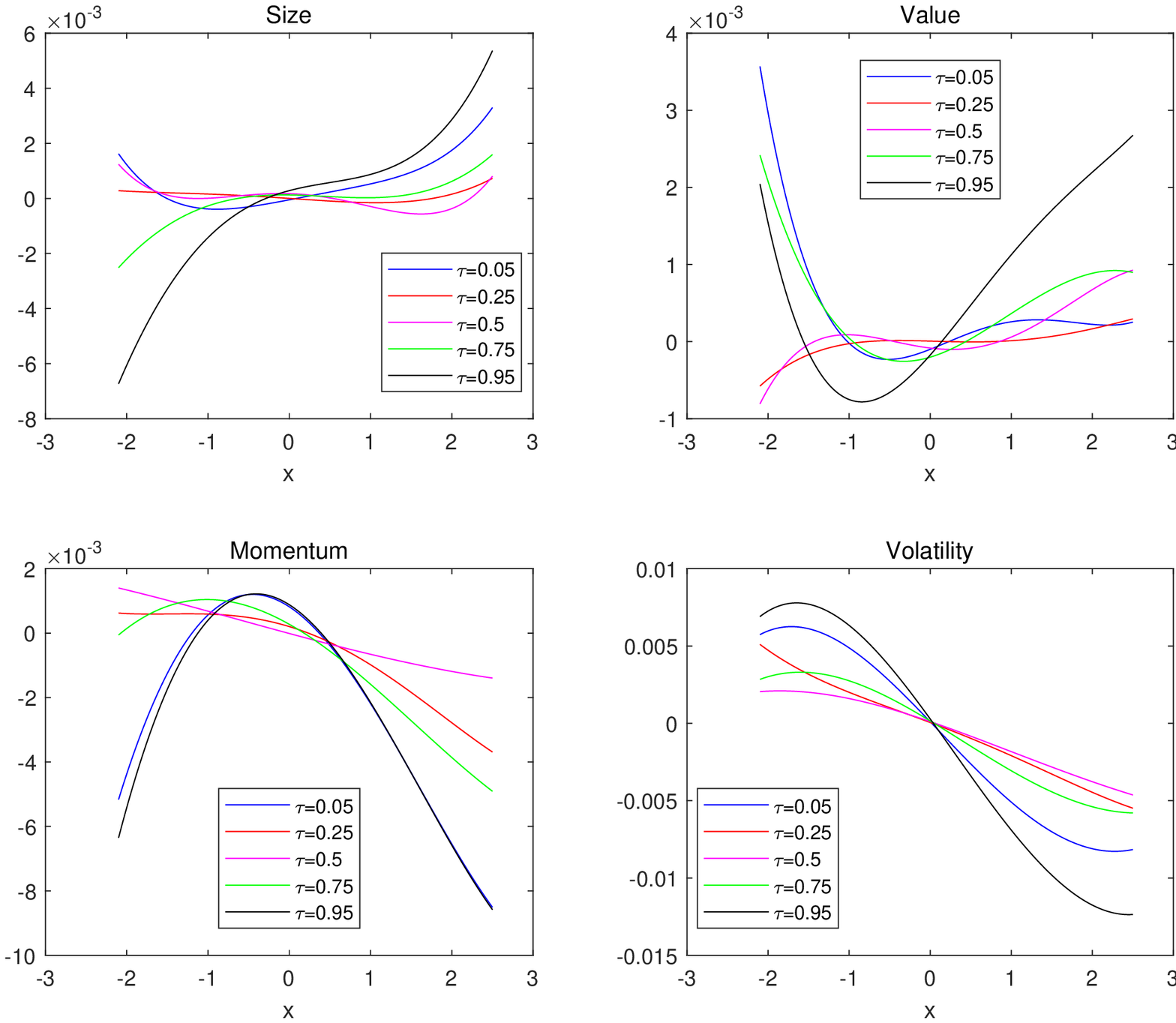}
			\captionsetup{singlelinecheck=off}
	\caption*{Note: This figure plots the estimated quantile factor loading functions of the four characteristics using QPPCA at different $\tau$s.}
		\end{figure}

 \section{Conclusions}
This paper proposes a three-stage estimation method for characteristic-based quantile factor models (CQFM). The convergence rates of the proposed estimators, labeled QPPCA, are established, and the asymptotic distributions of the estimated factors and loading functions are derived under very general conditions. Compared with the existing estimation methods of CQFM, not only QPPCA estimators are easier to implement in practice, but also they are consistent for fixed $T$ as long as $n$ goes to infinity, as well as being robust to heavy tails and outliers in the distribution of the idiosyncratic errors. Moreover, the number of quantile factors are allowed to be different from the number of the characteristics, and this number can be consistently estimated using a new rank-minimization estimator proposed in this paper.

An application of the estimators to a dataset consisting of individual stock returns reveals that the quantile factor loadings are nonlinear functions of some observed characteristics and that these functions exhibit considerable variations across quantiles. We conjecture that this leads to the concept of \textit{quantile returns} which generalizes the standard concept of \textit{expected returns}, typically proxied by averages of realized returns. 

The methodology associated with QPPCA is useful to derive the convergence rates and asymptotic properties of the (average)  quantile returns which remains high in our ongoing research agenda. Moreover, for the tractability of the problem, it has been assumed that the quantile factor loadings can be fully explained by the observed characteristics. Admittedly, this is a restrictive assumption. Relaxing this assumption and allowing the factor loadings to be functions of other unobserved characteristics is a challenging task in the context of quantile regressions. This interesting question is also left for future research.

 \newpage
\bibliographystyle{chicago}
\bibliography{QR_PCA}

\begin{thebibliography}{}

\bibitem[\protect\citeauthoryear{Ahn and Horenstein}{Ahn and
  Horenstein}{2013}]{ahn2013eigenvalue}
Ahn, S.~C. and A.~R. Horenstein (2013).
\newblock Eigenvalue ratio test for the number of factors.
\newblock {\em Econometrica\/}~{\em 81\/}(3), 1203--1227.

\bibitem[\protect\citeauthoryear{Bai}{Bai}{2003}]{bai2003inferential}
Bai, J. (2003).
\newblock Inferential theory for factor models of large dimensions.
\newblock {\em Econometrica\/}~{\em 71\/}(1), 135--171.

\bibitem[\protect\citeauthoryear{Bai and Ng}{Bai and
  Ng}{2002}]{bai2002determining}
Bai, J. and S.~Ng (2002).
\newblock Determining the number of factors in approximate factor models.
\newblock {\em Econometrica\/}~{\em 70\/}(1), 191--221.

\bibitem[\protect\citeauthoryear{Chamberlain and Rothschild}{Chamberlain and
  Rothschild}{1983}]{10.2307/1912275}
Chamberlain, G. and M.~Rothschild (1983).
\newblock Arbitrage, factor structure, and mean-variance analysis on large
  asset markets.
\newblock {\em Econometrica\/}~{\em 51\/}(5), 1281--1304.

\bibitem[\protect\citeauthoryear{Chen, Dolado, and Gonzalo}{Chen
  et~al.}{2021}]{chen2021quantile}
Chen, L., J.~J. Dolado, and J.~Gonzalo (2021).
\newblock Quantile factor models.
\newblock {\em Econometrica\/}~{\em 89\/}(2), 875--910.

\bibitem[\protect\citeauthoryear{Chen}{Chen}{2007}]{chen2007large}
Chen, X. (2007).
\newblock Large sample sieve estimation of semi-nonparametric models.
\newblock {\em Handbook of Econometrics\/}~{\em 6}, 5549--5632.

\bibitem[\protect\citeauthoryear{Chen and Shen}{Chen and
  Shen}{1998}]{chen1998sieve}
Chen, X. and X.~Shen (1998).
\newblock Sieve extremum estimates for weakly dependent data.
\newblock {\em Econometrica\/}, 289--314.

\bibitem[\protect\citeauthoryear{Connor, Hagmann, and Linton}{Connor
  et~al.}{2012}]{connor2012efficient}
Connor, G., M.~Hagmann, and O.~Linton (2012).
\newblock Efficient semiparametric estimation of the {F}ama--{F}rench model and
  extensions.
\newblock {\em Econometrica\/}~{\em 80\/}(2), 713--754.

\bibitem[\protect\citeauthoryear{Connor and Korajczyk}{Connor and
  Korajczyk}{1993}]{connor1993test}
Connor, G. and R.~A. Korajczyk (1993).
\newblock A test for the number of factors in an approximate factor model.
\newblock {\em the Journal of Finance\/}~{\em 48\/}(4), 1263--1291.

\bibitem[\protect\citeauthoryear{Connor and Linton}{Connor and
  Linton}{2007}]{connor2007semiparametric}
Connor, G. and O.~Linton (2007).
\newblock Semiparametric estimation of a characteristic-based factor model of
  common stock returns.
\newblock {\em Journal of Empirical Finance\/}~{\em 14\/}(5), 694--717.

\bibitem[\protect\citeauthoryear{Elton}{Elton}{1999}]{elton1999presidential}
Elton, E.~J. (1999).
\newblock Expected return, realized return, and asset pricing tests.
\newblock {\em The Journal of Finance\/}~{\em 54\/}(4), 1199--1220.

\bibitem[\protect\citeauthoryear{Fama and French}{Fama and
  French}{1993}]{fama1993common}
Fama, E.~F. and K.~R. French (1993).
\newblock Common risk factors in the returns on stocks and bonds.
\newblock {\em Journal of Financial Economics\/}~{\em 33\/}(1), 3--56.

\bibitem[\protect\citeauthoryear{Fama and French}{Fama and
  French}{2015}]{fama2015five}
Fama, E.~F. and K.~R. French (2015).
\newblock A five-factor asset pricing model.
\newblock {\em Journal of Financial Economics\/}~{\em 116\/}(1), 1--22.

\bibitem[\protect\citeauthoryear{Fan, Li, and Liao}{Fan
  et~al.}{2021}]{fan2021recent}
Fan, J., K.~Li, and Y.~Liao (2021).
\newblock Recent developments in factor models and applications in econometric
  learning.
\newblock {\em Annual Review of Financial Economics\/}~{\em 13}, 401--430.

\bibitem[\protect\citeauthoryear{Fan, Liao, and Wang}{Fan
  et~al.}{2016}]{fan2016projected}
Fan, J., Y.~Liao, and W.~Wang (2016).
\newblock Projected principal component analysis in factor models.
\newblock {\em Annals of Statistics\/}~{\em 44\/}(1), 219.

\bibitem[\protect\citeauthoryear{Hausman, Liu, Luo, and Palmer}{Hausman
  et~al.}{2021}]{hausman2021errors}
Hausman, J., H.~Liu, Y.~Luo, and C.~Palmer (2021).
\newblock Errors in the dependent variable of quantile regression models.
\newblock {\em Econometrica\/}~{\em 89\/}(2), 849--873.

\bibitem[\protect\citeauthoryear{Horowitz and Lee}{Horowitz and
  Lee}{2005}]{horowitz2005nonparametric}
Horowitz, J.~L. and S.~Lee (2005).
\newblock Nonparametric estimation of an additive quantile regression model.
\newblock {\em Journal of the American Statistical Association\/}~{\em
  100\/}(472), 1238--1249.

\bibitem[\protect\citeauthoryear{Jung and Marron}{Jung and
  Marron}{2009}]{jung2009pca}
Jung, S. and J.~S. Marron (2009).
\newblock Pca consistency in high dimension, low sample size context.
\newblock {\em The Annals of Statistics\/}~{\em 37\/}(6B), 4104--4130.

\bibitem[\protect\citeauthoryear{Kato, Galvao, and Montes-Rojas}{Kato
  et~al.}{2012}]{kato2012asymptotics}
Kato, K., A.~F. Galvao, and G.~V. Montes-Rojas (2012).
\newblock Asymptotics for panel quantile regression models with individual
  effects.
\newblock {\em Journal of Econometrics\/}~{\em 170\/}(1), 76--91.

\bibitem[\protect\citeauthoryear{Lee and Robinson}{Lee and
  Robinson}{2016}]{lee2016series}
Lee, J. and P.~M. Robinson (2016).
\newblock Series estimation under cross-sectional dependence.
\newblock {\em Journal of Econometrics\/}~{\em 190\/}(1), 1--17.

\bibitem[\protect\citeauthoryear{Ma, Linton, and Gao}{Ma
  et~al.}{2021}]{ma2021estimation}
Ma, S., O.~Linton, and J.~Gao (2021).
\newblock Estimation and inference in semiparametric quantile factor models.
\newblock {\em Journal of Econometrics\/}~{\em 222\/}(1), 295--323.

\bibitem[\protect\citeauthoryear{Shen, Shen, and Marron}{Shen
  et~al.}{2013}]{shen2013consistency}
Shen, D., H.~Shen, and J.~S. Marron (2013).
\newblock Consistency of sparse pca in high dimension, low sample size
  contexts.
\newblock {\em Journal of Multivariate Analysis\/}~{\em 115}, 317--333.

\bibitem[\protect\citeauthoryear{Stock and Watson}{Stock and
  Watson}{2002}]{stock2002forecasting}
Stock, J.~H. and M.~W. Watson (2002).
\newblock Forecasting using principal components from a large number of
  predictors.
\newblock {\em Journal of the American Statistical Association\/}~{\em
  97\/}(460), 1167--1179.

\end{thebibliography}

\end{document}